\documentclass{ws-procs9x6}

\def\ls{{_<\atop^{\sim}}}
\def\gs{{_>\atop^{\sim}}}
\def\cgs{ ${\rm erg~cm}^{-2}~{\rm s}^{-1}$ }

\begin{document}

\title{High Energy Large Area Surveys: 
optically obscured AGN and the history of accretion
\footnote{\uppercase{T}his work is partially supported by \uppercase{ASI}
grant I/R/057/02, by \uppercase{INAF} grant \# 270/2003 and \uppercase{MIUR} 
grant Cofin--03--02--23}}

\author{F. Fiore$^1$ and the HELLAS2XMM collaboration}

\address{$^1$ INAF-Osservatorio Astronomico di Roma\\
via Frascati, 33, I00040, Monteporzio, Italy\\ 
E-mail: fiore@mporzio.astro.it}


\maketitle

\abstracts{Hard X-ray, large area surveys are a fundamental complement
of ultra-deep, pencil beam surveys in obtaining a more complete
coverage of the L--z plane, allowing to find luminous QSO in 
wide z ranges. Furthermore, results from these surveys can be
used to make reliable predictions about the luminosity (and
hence the redshift) of the sources in the deep surveys which have
optical counterparts too faint to be observed with the present
generation of optical telescopes. This allows us to obtain
accurate luminosity functions on wide luminosity and redshift
intervals.
}

\section{Introduction}

Hard X-ray surveys are the most direct probe of super-massive black
hole accretion activity, which is recorded in the cosmic X-ray
background (CXB) spectral intensity.  Deep, pencil beam, Chandra and
XMM-Newton surveys have resolved most of the CXB below 4-6 keV, and
$\approx50\%$ of the CXB at 10 keV (Worsley et al. 2004).  The optical
spectroscopic follow-up of the Chandra Deep Field North/South (CDFN,
CDFS) and of the Lockman Hole (LH) surveys proved to be very efficient
at identifying a large population of Seyfert-like objects up to
z=2--3, and a few QSOs up to z=4, see Fig. \ref{surveys}a and
references in Table 1). Shallower, but larger area surveys are
therefore fundamental to: a) complement the pencil-beam survey to
obtain a more complete coverage of the L--z plane, to find QSO with
logL(2-10keV)$>$44, i.e. close to  AGN luminosity
function L$^*$, in a large z range; b) to obtain reliable spectroscopic
redshifts of the faint optical counterpats of sources with X-ray to
optical flux ratio (X/O) much higher than that of typical broad line
AGN (X/O$\sim1$), which make 20--30\% of the full samples. At the
10-100 times fainter fluxes reached by the CDFN, CDFS and LH surveys
most sources with high X/O have optical counterparts too faint for
even 8--10 m class telescopes.  The best hope to obtain information on
this elusive X-ray source population is to make use of the information
gained at higher fluxes to make reliable predictions about their
luminosities and redshifts.

\subsection{Source samples}

\begin{figure}[ht]
\centerline{
\hbox{
\epsfxsize=2.3in\epsfbox{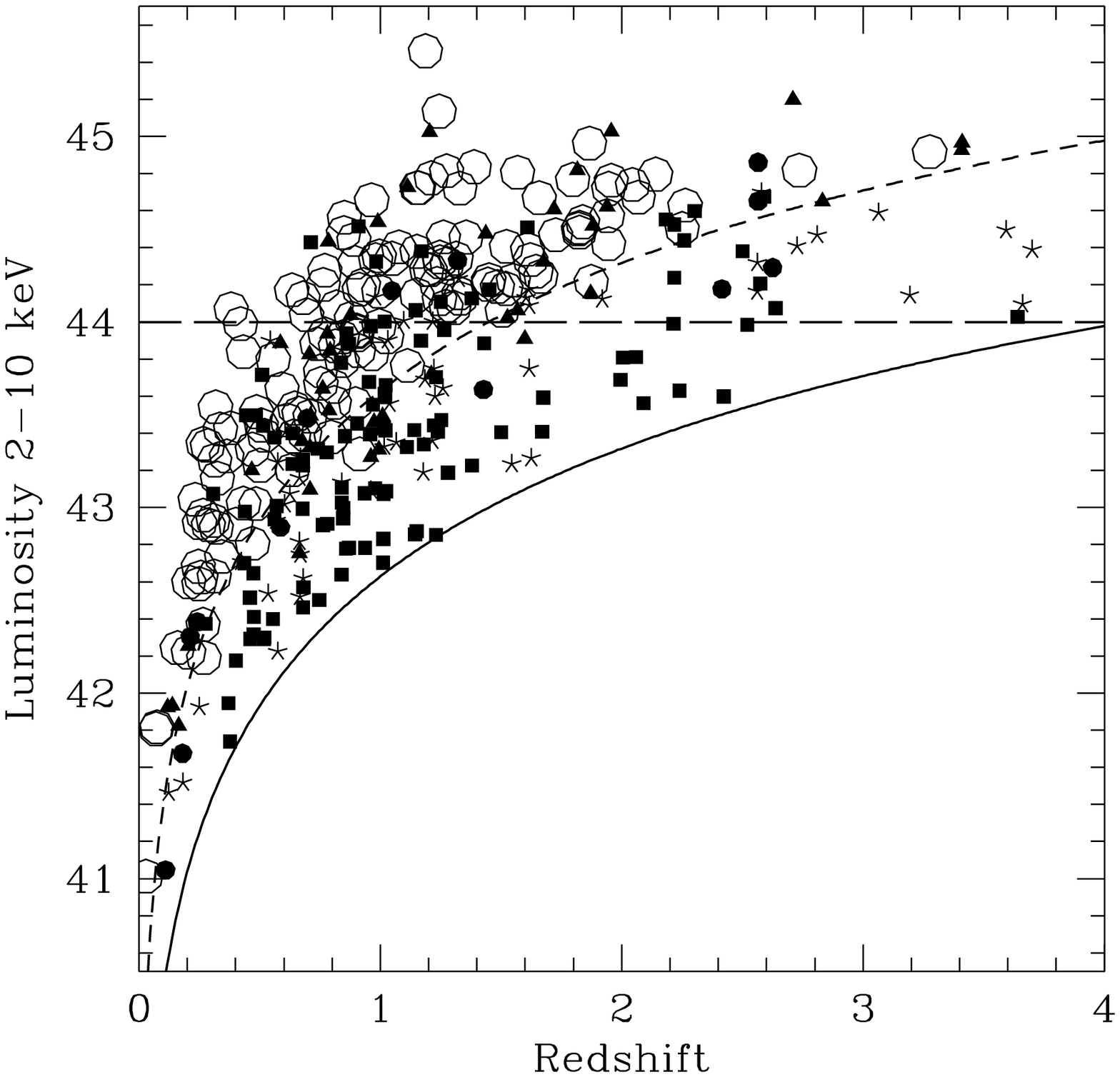}
\epsfxsize=2.3in\epsfbox{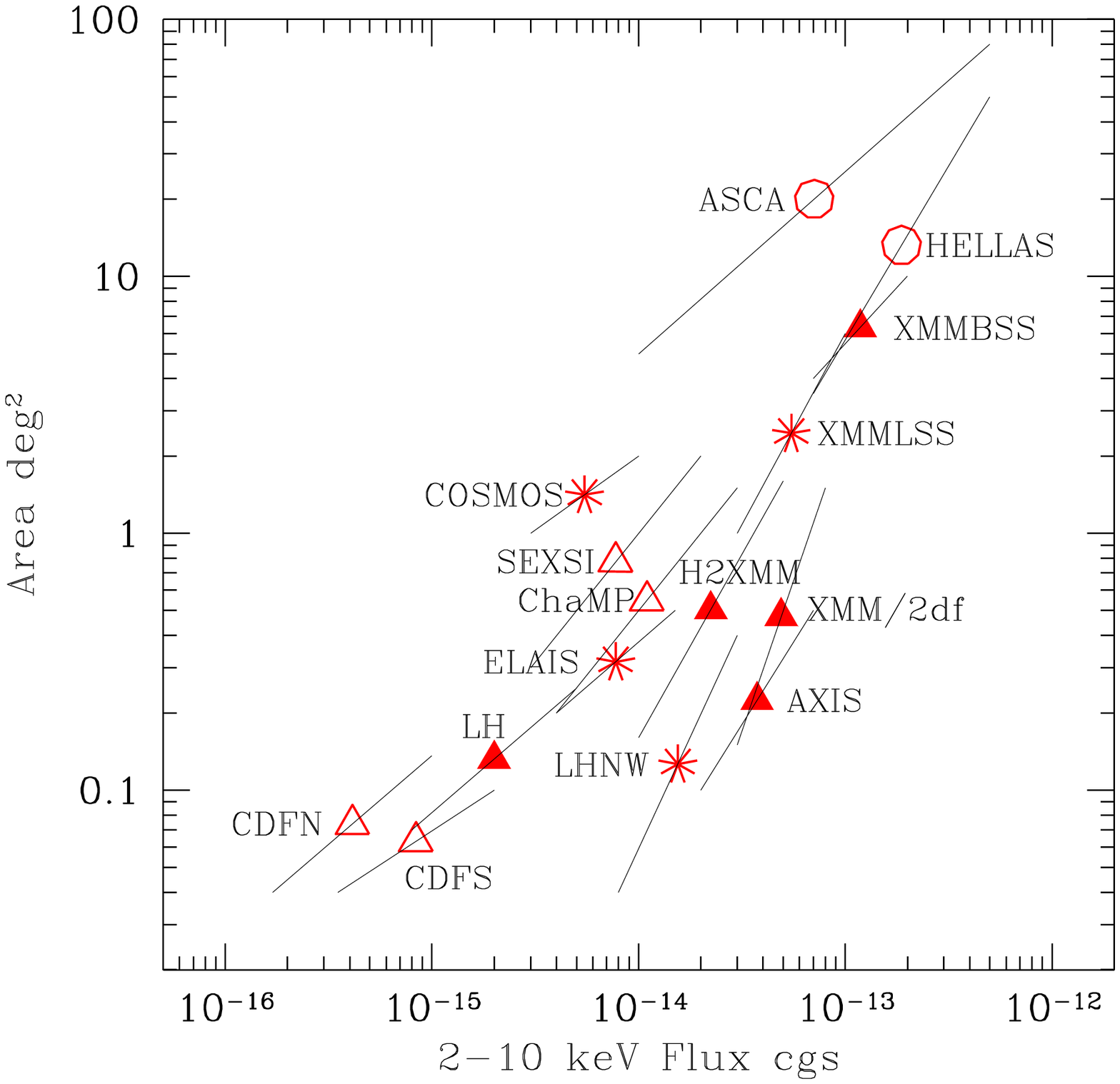}
}   
}
\caption{
\footnotesize{
Left: the L(2-10keV)--z plane for the source samples in Table
1 (symbols as in Table 1).  The lower (upper) solid (dashed) lines
represent the flux limits of $10^{-15}$ \cgs (i.e. Chandra deep
surveys) and $10^{-14}$ \cgs (i.e.  HELLAS2XMM survey)
respectively. Right: the flux-area diagram for several 2-10 keV
surveys.  Open triangles =  Chandra surveys, filled triangles = XMM-Newton 
surveys, stars = Chandra or XMM-Newton ``contiguous'' mosaics}
}
\label{surveys}
\end{figure}

Figure \ref{surveys}b) compares the fluxes and area covered by a
number of hard X-ray surveys.  In this paper we use the source samples
given in Table 1, which include the deepest surveys performed with
Chandra and the shallower, but much larger area HELLAS2XMM survey. As
of today we have obtained spectroscopic redshifts for more than 150
sources, and, most important, we were able to obtain spectroscopic
redshifts and classification of many sources with X/O$>10$; finding
about ten type 2 QSO at z=0.7--2, to be compared with the similar
number obtained from the combination of the CDFN and CDFS, at the
expenses of a huge investment of VLT and Keck observing time. For
other 8 X/O$>10$ sources a redshift estimate was obtained from their
observed R--K colors (Mignoli et al. 2004).

\begin{table}[ph]
\caption{\bf 2-10 keV surveys}
{\footnotesize
\begin{tabular}{lcccccc}
\hline
Sample     & Tot. Area & Flux limit & \# sour. & \% z-spec & refs & symbol\\
           & deg$^2$   & $10^{-15}$ cgs &       &           &      &\\
\hline
HELLAS2XMM      & 1.6   & 8.0       &  231      & 66\%      & 1,2  & open circ.\\
CDFN faint$^a$  & 0.0369& 1.0       &   88      & 59\%      & 3,4  & filled squar.\\ 
CDFN bright$^b$ & 0.0504& 3.0       &   44      & 65\%      & 3,4  & filled squar.\\
CDFS faint$^a$  & 0.0369& 1.0       &   68      & 62\%      & 5,6  & stars  \\
CDFS bright$^b$ & 0.0504& 3.0       &   55      & 58\%      & 5,6  & stars  \\ 
Lockman Hole$^c$& 0.126 & 4.0       &   55      & 75\%      & 7    & filled triang.\\
SSA13$^d$       & 0.0177& 3.8       &   20      & 65\%      & 8    & filled circ.\\
\hline
Total           &       &           &  561      & 65\%      &      & \\
\hline
\end{tabular}
\begin{tabnote}
$^a$ Inner 6.5 arcmin radius; $^b$ outer 6.5--10 arcmin annulus; $^c$ inner 12 arcmin
radius; $^d$ inner 4.5 arcmin radius; (1) Fiore et al. 2003; (2) Cocchia et al. in 
prep.; (3) Alexander et al. 2003; (4) Barger et al. 2003; (5) Giacconi et al.
2002; (6) Szokoly et al. 2004; (7) Mainieri et al. 2002; (8) Barger et al. 2001. 
\end{tabnote}
}
\label{table1} 
\vspace*{-13pt}
\end{table}

\section{Optically obscured AGN: a robust method to estimate 
their luminosity}

\begin{figure}[ht]
\centerline{
\hbox{
\epsfxsize=2.1in\epsfbox{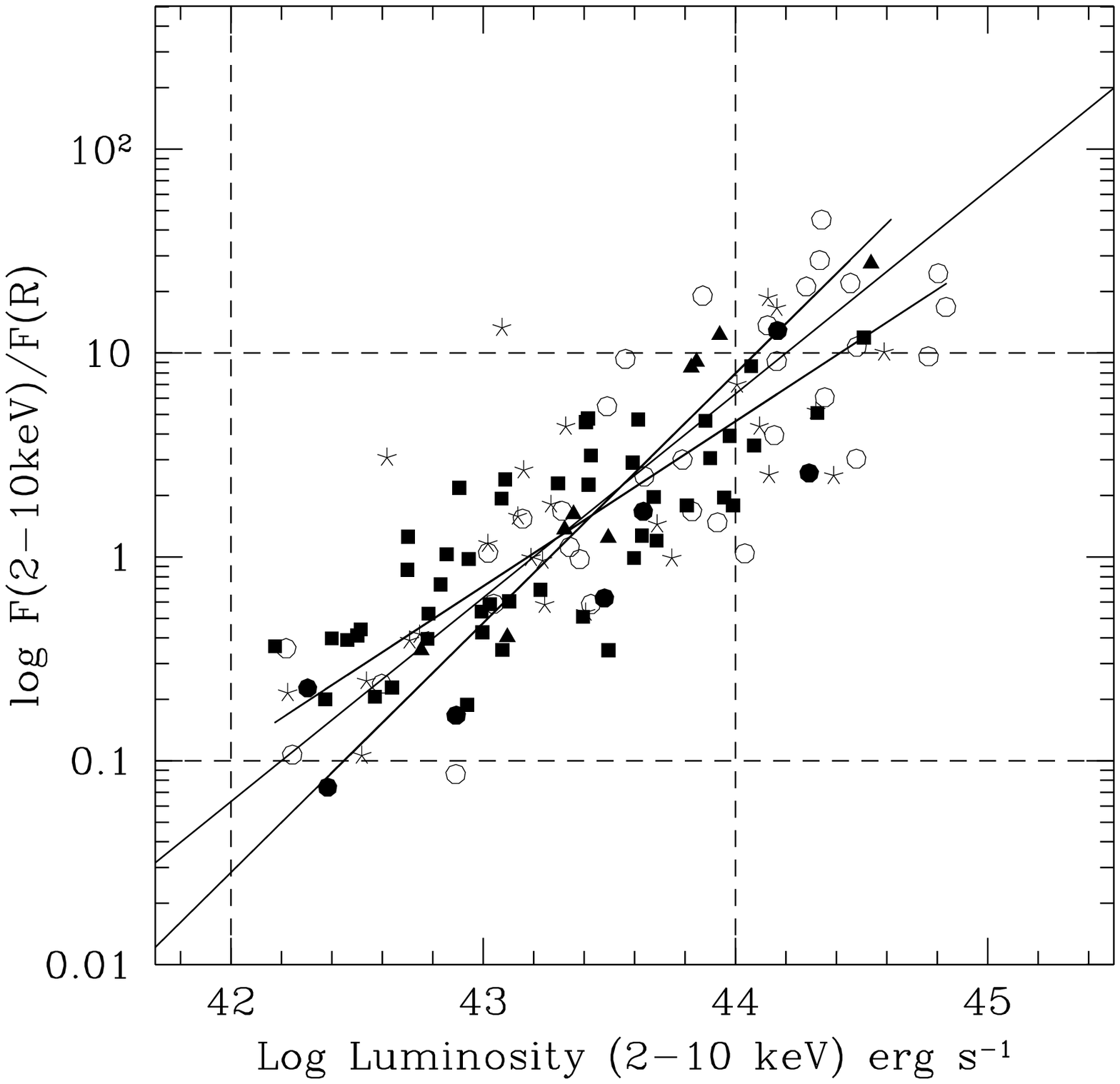}
\epsfxsize=2.1in\epsfbox{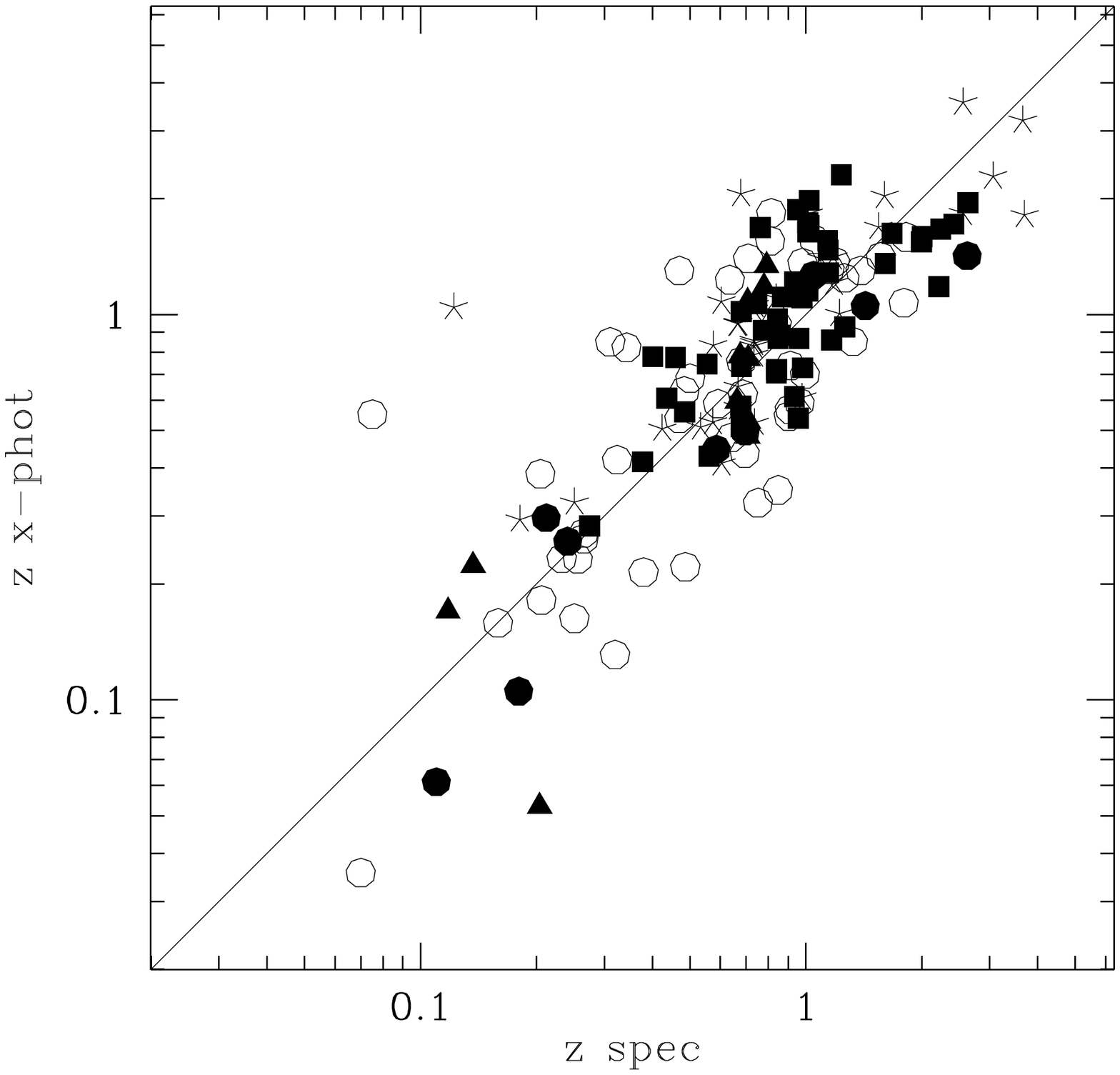}
}
}
\caption{
\footnotesize{Left: the X-ray to optical ratio as a function of the X-ray luminosity 
for 123 optically obscured AGN with secure redshift and classification from
the samples in Table 1. The three solid lines are the least squares fit 
of y=y(x) and x=x(y) and their average.
Right: spectroscopic redshift versus the redshifts
evaluated using the correlation in the left panel and the X-ray flux.
}}
\label{zz}
\end{figure}

Fiore et al. (2003) discovered a striking correlation between X/O and
the 2--10 keV luminosity for the sources with the nucleus strongly
obscured in the optical band, i.e. not showing broad emission lines
(Fig. \ref{zz}a).
The dispersion along the correlation in Fig. \ref{zz}a) is of 0.40 dex
in luminosity.  Note as at X/O$\gs10$ most of the sources are from the
HELLAS2XMM sample, i.e. it is mostly thank to this sample that it is
possible to extend and validate the correlation at high X/O values,
and therefore at high luminosities.  From X/O, and hence the
luminosity, it is possible to estimate the redshift of optically
obscured sources. We call the resulting redshift estimates
``X-photo-z''.  Fig. \ref{zz}b) plots our X-photo-z against the
spectroscopic redshifts.  The correlation is again rather good and
$\sigma(\Delta z/(1+z)\sim0.2)$. This can be compared with the value
of $\sim0.1$, typical of accurate ``photometric'' redshift estimates.
We note that in our case only 2 bands are used, agaist the several
O-NIR bands necessary to obtain reliable photo-z. Our present
X-photo-z estimates are obtained using a linear fit to the
logX/O-logL(2-10keV) relationship.  More accurate estimates can be
obtained using higher order polinomia (Fiore et al. 2004 in prep.).

\subsection{Optical vs X-ray obscuration: the ``Figure of Merit''}

\begin{figure}[ht]
\centerline{
\hbox{
\epsfxsize=2.0in\epsfbox{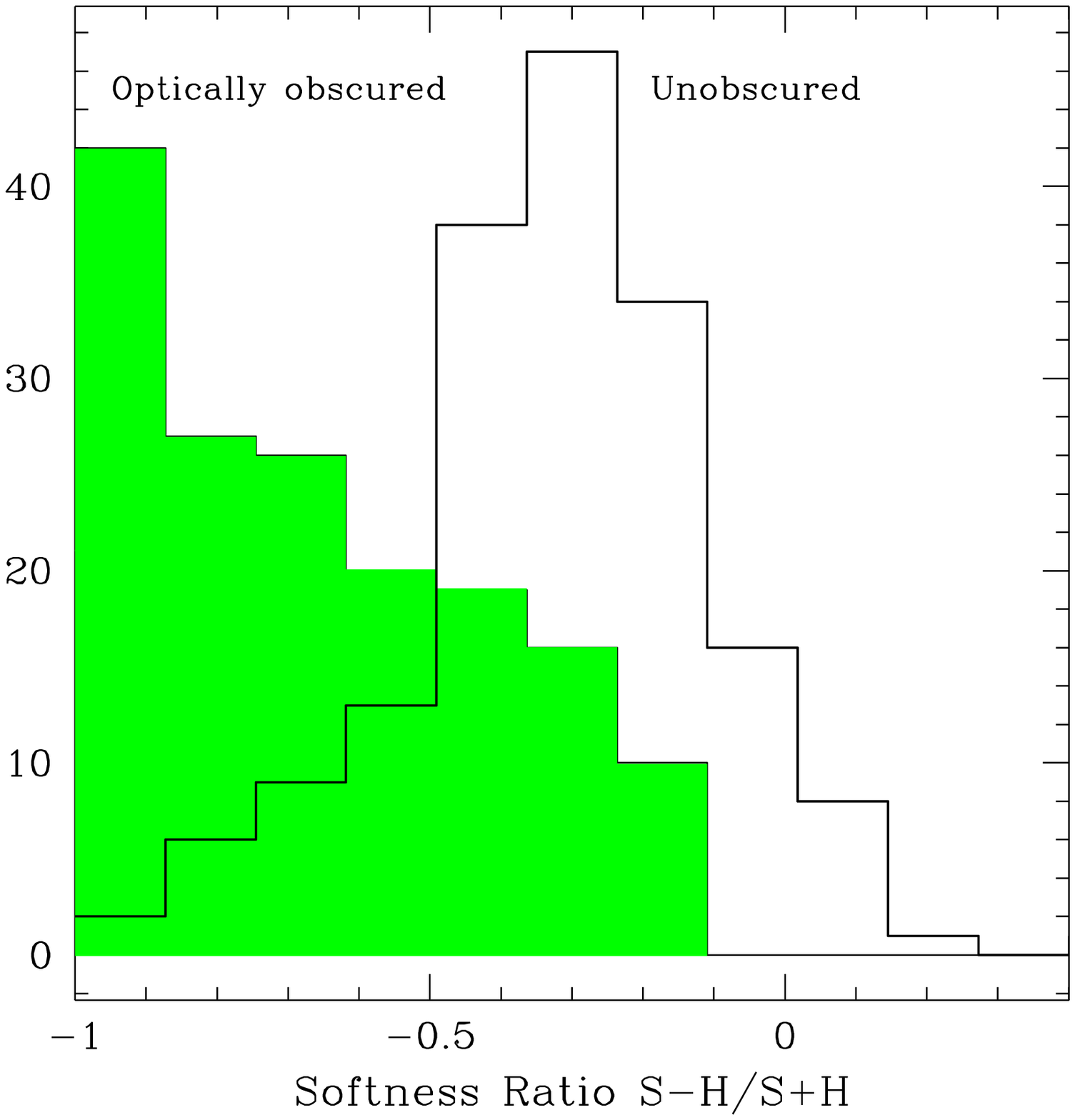}
\epsfxsize=2.0in\epsfbox{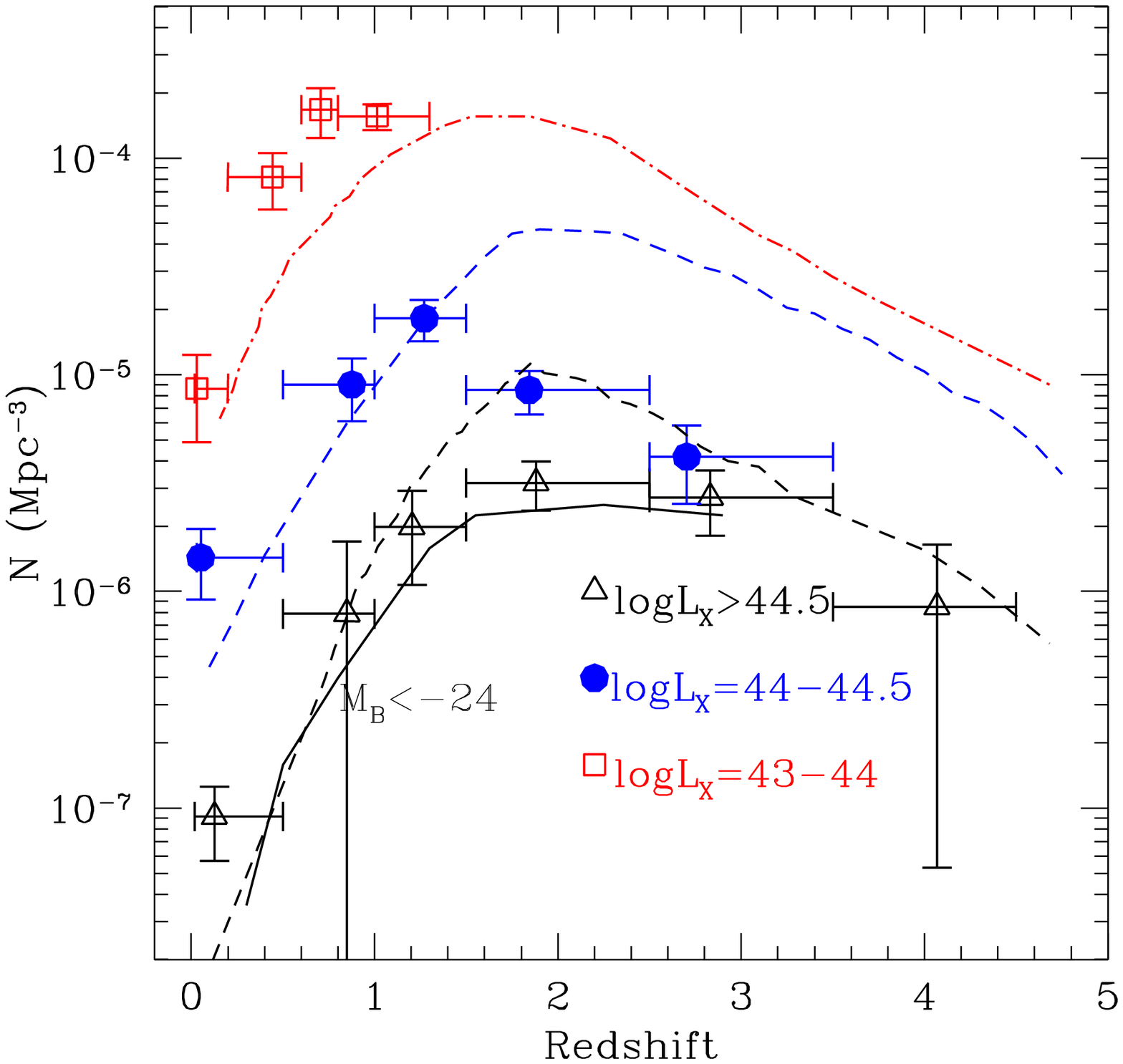}
}
}   
\caption{ 
{\footnotesize
Left: the distribution of the softness ratio of optically
obscured and unobscured sources, Right: the evolution of the number
density of hard X-ray selected AGNs in three luminosity bins: The
solid curve represents the evolution of optically selected QSO
with M$_B\ls-24$.  The three dashed curves are predictions from Menci
et al. 2004}
}
\label{hrt}
\end{figure}

Perola et al. (2004) found that most type 2 AGN in the HELLAS2XMM
sample have significant obscuration (rest frame absorbing column
$N_H>10^{22}$ cm$^{-2}$) also in the X-rays.  Perola et al. (2004) and
Comastri \& Fiore (2004) also found a good correlation between X/O and
$N_H$.  Both facts suggest that X-ray sources log$N_H>22$ are likely
to be also optically obscured AGN. However, for many weak sources in
the samples in Table 1 $N_H$ proper spectral fits are unfeasible and a
rough spectral information can be derived from their softness ratio
only. The correlation between the softness ratio and $N_H$ is
reasonably good (Perola et al. 2004) and therefore using the former to
select X-ray obscured AGN is not a bad approximation.  The
relationship between X-ray obscured AGN and optically obscured AGN is
not one to one. This is clearly seen in Fig. \ref{hrt}b), which shows
that $\sim90\%$ of type 1 AGN have (S-H)/(S+H)$>-0.5$ (i.e. most of
them have a soft spectrum) and $\sim70\%$ of optically obscured AGN
have (S-H)/(S+H)$<-0.5$ (i.e. the majority have a hard, likely
obscured spectrum, but $\sim30\%$ have a soft spectrum).  This means
that using a single (S-H)/(S+H) threshold to decide whether an X-ray
source with a faint optical counterpart is an optically obscured AGN
has an intrinsically large uncertainty.  For this reason, we instead
compute for each source a ``figure of merit'' (FoM, Fiore et al. 2004
in prep.), using: 1) the X/O ratio and the fraction of
obscured/unobscured sources as a function of X/O (see Fiore et
al. 2003); 2) the morphology of the optical counterpart, i.e
point-like sources are likely to be optically unobscured sources; 3)
the probability distribution of the softness ratio for obscured and
unobscured AGN, as estimated for the sources with spectroscopic
redshift and classification in Table 1. If a source has a FoM
qualifying it as optically obscured we use the correlation in
fig. \ref {zz}a) to estimate its redshift; if a source qualifies as an
optically unobscured AGN we guess its redshift using the loose
relation between X-ray flux and redshift for type 1 AGN. These
estimates should therefore be considered in a statistical sense only
(Fiore et al. 2003).  At fluxes lower than $10^{-14}$ \cgs only
$\approx30\%$ of the sources qualify as unobscured AGN.

\section{The evolution of hard X-ray selected sources}

Once redshifts and luminosities of the sources of the full sample are
known (either spectroscopically measured or photometrically and
statistically estimated) we can compute the luminosity function of
hard X-ray selected sources and its redshift
evolution. Fig.\ref{hrt}b) plots the AGN number density in three
luminosity bins as a function of z.  This is compared with
the number density of luminous optically selected QSOs and with the
prediction of the Menci et al. (2004) semi-analytic, hierarchical
clustering model (MM). While the density of luminous AGN increases
monotonically up to z$\sim3$, following the evolution of optically
selected AGN, and in agreement with the MM, the evolution of
increasingly lower luminosity AGN peaks at decreasingly lower
redshifts. The peak redshifts are also lower than those of the MM
predictions. At z$\gs1-2$ the density of logL=$44-44.5$ AGN decreases
sharply, and it is inconsistent with the MM predictions, while the
density of logL$<44$ AGN is not well constrained by the present
data. An extenction of the present analysis to flux limits lower than
those in Table 1, and therefore to higher z, is in progress (Fiore et
al. 2004 in prep.).  The paucity of Seyfert-like objects at z$\gs1-2$
can be due to at least two reasons: a) a selection effect, i.e.
highly obscured AGN are common at these redshifts (as in the nearby
Universe) but are missed (or their luminosity is underestimated) in
Chandra and XMM-Newton surveys (La Franca et al. 2004 in prep,; b) a
different description of the mechanisms regulating the amount of cool
gas in low-mass host galaxies, the physical mechanism at work at small
accretion rates and/or the statistics of DM condensations is needed in
the MM.  To avoid possible selection effect, for an unbiased census of
the AGN population making the bulk of the CXB and an unbiased measure
of the AGN luminosity function at z=1--2, the "golden epoch'' of galaxy
and AGN activity, sensitive observations extending at energies where
photoelectic effect no more reduces the observed flux (i.e. E$>10$ keV)
are clearly needed.  More specifically to resolve $\sim$50\% of the
20--100 keV CXB we need to go down to fluxes of $10^{-14}$ \cgs in
this band (see fig. 6 of Menci et al. 2004).  This can be achieved
only by imaging X-ray telescopes, possibly with multi--layer coatings
(see e.g. Pareschi \& Cotroneo 2003).  Key issues are: a) high
collecting area; b) sharp PSF (15 arcsec or less Half Energy Diameter,
HED); c) low detector internal background.

\section*{Acknowledgments}
{\footnotesize
The original matter presented in this paper is the
result of the effort of a large number of people, in particular of the
{\tt HELLAS2XMM} team (A. Baldi, 
M. Brusa, F. Cocchia, N. Carangelo, P. Ciliegi, F. Cocchia, A. Comastri,
V. D'Elia, C. Feruglio, F. La Franca, R. Maiolino, G. Matt, M. Mignoli,
S. Molendi, G. C. Perola, S. Puccetti, C. Vignali), N. Menci, A. Cavaliere,
M. Elvis and P. Severgnini.
}

\end{document}